\documentclass[aps,prb,amsmath,amssymb,preprintz,twocolumn]{revtex4-2}

\usepackage{graphicx}
\usepackage{amsmath}
\usepackage{braket}
\usepackage{amssymb}
\usepackage{bbm}
\usepackage{bm}
\usepackage{mathtools}
\usepackage{hyperref}
\usepackage{makecell}
\usepackage{booktabs}
\usepackage{siunitx}
\usepackage{natbib}
\usepackage{mathtools}\usepackage[normalem]{ulem}
\graphicspath{{Figures}}
\usepackage[normalem]{ulem}

\usepackage[x11names]{xcolor}

\newcommand{\Reff}{R^{\rm eff}}
\newcommand{\mmax}{\ensuremath{m_{\rm max}}}
\newcommand{\hthres}{h_{\rm thres}}

\begin{document}
	
	\title{Societal self-regulation induces complex infection dynamics and chaos}

    \author{Joel Wagner}
    \altaffiliation{JW and SB contributed equally.}
    
    \author{Simon Bauer}
    \altaffiliation{JW and SB contributed equally.}
    
    \author{Sebastian Contreras}
    
    \author{Luk Fleddermann}
    
    \author{Ulrich Parlitz}
    
    \author{Viola Priesemann}
    \altaffiliation{Corresponding Author: Viola Priesemann\\ (viola.priesemann@ds.mpg.de)}
    \affiliation{Max Planck Institute for Dynamics and Self-Organization, G\"ottingen, Germany.}
    \affiliation{Institute for the Dynamics of Complex Systems, University of G\"ottingen,  G\"ottingen, Germany.}
    
    \date{\today}

	\begin{abstract}
    Classically, endemic infectious diseases are expected to display relatively stable, predictable infection dynamics.
    Accordingly, basic disease models such as the susceptible-infected-recovered-susceptible model display stable endemic states or recurrent seasonal waves. 
    However, if the human population reacts to high infection numbers by mitigating the spread of the disease, then this delayed behavioral feedback loop can generate infection waves itself, driven by periodic mitigation and subsequent relaxation.
    We show that such behavioral reactions, together with a seasonal effect of comparable impact, can cause complex and unpredictable infection dynamics, including Arnold tongues, co-existing attractors, and chaos. Importantly, these arise in epidemiologically relevant parameter regions where the costs associated to infections and mitigation are jointly minimized. 
    By comparing our model to data, we find signs that COVID-19 was mitigated in a way that favored complex infection dynamics.
    Our results challenge the intuition that endemic disease dynamics necessarily implies predictability and seasonal waves, and show the emergence of complex infection dynamics when humans optimize their reaction to increasing infection numbers.

\end{abstract}

	\maketitle

\section{Introduction}

Infectious diseases have always accompanied human societies. Some of them pose a major threat and require mitigation measures. If infection waves can be well anticipated, for example if they clearly follow a seasonal pattern, this allows to plan such mitigation measures. However, human mitigation itself adds another uncertainty to disease spread, further challenging reliable forecasts, and ultimately adequate policies.

Infection dynamics is classically studied in agent-based or compartmental susceptible-infected-recovered (-susceptible) [SIR(S)] like models \cite{kermack1927contribution,wang2016statistical,manfredi2013modeling}. In the latter, the population is split into disjoint compartments according to their disease status. Susceptible individuals (S) get infected by infectious individuals (I), recover (R) and stay immune until they eventually become susceptible again (S). In these models, infection dynamics is well predictable, as it settles into an endemic equilibrium, possibly modulated yearly by seasonality. However, it has been suggested that an interplay of seasonality and human behavior leads to chaotic dynamics around this equilibrium \cite{d2009information}. Here, we investigate this systematically by analyzing a SIRS model including mitigation and seasonality as a driven nonlinear oscillator.
In dynamical systems theory, such driven oscillators are known to display complex steady-state dynamics \cite{parlitz1987period,Mettin_et_al_1993,datseris2022nonlinear}, including regimes of phase-locking, chaos and coexisting attractors. Here, we find such highly complex disease states in epidemiologically relevant parameter regions.

\section{Model Overview}

\begin{figure}
    \centering
    \includegraphics[width=8.6cm]{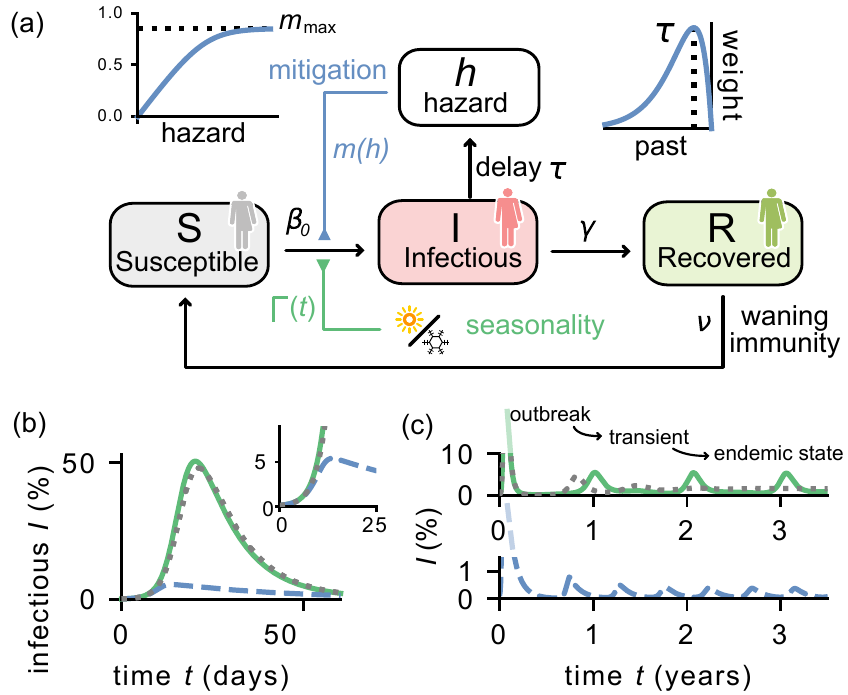}
    \caption{ 
        \textbf{Model overview} (a) The extended SIRS model including periodic seasonality ($\Gamma$, green) and delayed behavioral feedback for mitigation ($m$, blue). 
        (b) For a novel disease outbreak, classic SIRS models (with or without seasonality $\Gamma$, resp. green and dotted gray) feature initial exponential growth of infection numbers, eventually stopped only by immunity. With behavioral feedback, mitigation keeps the infection number considerably lower (dashed blue, see inset).
        (c) Following a transient phase, infection numbers settle into their endemic state. In the SIRS model with neither seasonality nor feedback, this endemic state is stable (dotted gray). In contrast, both seasonality (green) and behavioral feedback (dashed blue) can induce oscillations on their own. 
        Parameters: $\beta_0=0.5, \gamma=0.1$. Green: $a=0.25$, $\mmax=0$, $\nu=1/500$. Dashed blue: $a=0$, $\mmax=0.84$, $\tau=30$, $\nu=1/100$.
        }
    \label{fig:1-model-overview}
\end{figure}

There are various ways to incorporate human behavior into disease models \cite{funk2010modelling,d2009information,manfredi2013modeling,doenges2022interplay,bauch2013behavioral}.
One common \textit{behavior-implicit} approach, i.e., including human behavior without explicitly modeling human actions, is to assume that high infection numbers are perceived as increased \textit{hazard} $h(t)$, leading to stronger mitigation $m(h)$. The perceived hazard is quantified by the convolution of past infections with a weighting kernel $K_\tau$ and characteristic mitigation delay $\tau$ \cite{wang2016statistical,d2007vaccinating,d2009information}:

\begin{equation}
    h(t) = \int \limits_{-\infty}^t I(t')\, K_\tau(t-t')\, \text{d}t'
\end{equation}

with $K(t)=\frac{1}{\tau^2}te^{-t/\tau }$, that peaks at $\tau$ days in the past (Fig.~\ref{fig:1-model-overview}a). The mitigation delay $\tau$ embodies various delays affecting disease mitigation, such as the reporting of cases or the accumulation of risk perception over time. Mitigation includes various measures to slow the spread of the disease, such as reducing contacts, quarantining, mask-wearing, and other health-protective behaviors. These measures may be adopted voluntarily or mandated by the government, both typically occurring in response to high infection numbers and thus perceived hazard.

The hazard $h(t)$ causes mitigation $m(h)$, reducing the effective reproduction number $\Reff$ by a factor (1-$m(h))$. We assume that mitigation $m(h)$ i) increases monotonously with hazard $h$, and ii) saturates at a level $\mmax$ (Fig.~\ref{fig:1-model-overview}a, Tab.~\ref{tab:compartmentsfunctions}). This saturation encodes that people cannot reduce the infection rate to zero. The precise choice of the shape of $m(h)$ does not impact our results (Sec.~\ref{sec:s-sensitivity}). 
Modeling seasonality as a periodic multiplicative driving $\Gamma(t)=1+a\cos(\omega t)$, the number of susceptible ($S$), infectious ($I$) and recovered ($R$) individuals follows 

\begin{align}
    \dot  S &=  -\gamma R_0 \left(1-m(h)\right)\Gamma(t)\,IS +\nu R \label{eq:modelmaintext1}\,,\\
    \dot  I &= \hspace{0.3cm} \gamma R_0 \left(1-m(h)\right)\Gamma(t)\,IS - \gamma I \label{eq:modelmaintext2}\,,
\end{align}
and $R=1-S-I$.

The basic reproduction number $R_0=\frac{\beta_0}{\gamma}$ is expressed as the ratio between the basic reproduction rate $\beta_0$ and recovery rate $\gamma$. In the following we consider a disease with $R_0=5$, fast waning of immunity $\nu=1/100$ and a steep increase of $m(h)$, which could be interpreted as a highly infectious and lethal respiratory disease (Tab.~\ref{tab:compartmentsfunctions} for all variables and default values, Sec.~\ref{sec:s-sensitivity} for results with different parameters).

In comparison to the classic SIRS model, behavioral mitigation markedly changes the infection dynamics: Increasing hazard $h$ triggers mitigation, and thus exponential growth is broken earlier, which reduces the height of the wave (Fig.~\ref{fig:1-model-overview}b).

In the endemic state, which we here define as the dynamic steady state after the outbreak and transient phase, the infection dynamics in the classic SIRS model settles into a stable endemic equilibrium. Seasonality modulates this equilibrium into seasonal infection waves (Fig.~\ref{fig:1-model-overview}c, green). 
However, if a disease outbreak is so strong that mitigation becomes necessary, oscillations can emerge, even without seasonality (Fig.~\ref{fig:1-model-overview}c, dashed blue). These are purely a result of the delayed behavioral feedback and emerge through a Hopf bifurcation \cite{d2009information,liu2015endemic}, if the maximal achievable mitigation $\mmax$ is strong enough to break waves (i.e., $(1-\mmax)R_0 < 1$), and the mitigation delay $\tau$ is long enough, so that the wave has time to build up. This also becomes apparent in the stability diagram of the endemic state, where the Hopf bifurcation curve separates the stable regime from the regime with oscillations (Fig.~\ref{fig:2-temporal-simulations}a, Sec.~\ref{sec:s-stability-ee} for details on the computation). Very short delay $\tau$ or insufficient maximal mitigation $\mmax$ merely reduce the effective $\Reff$, and hence do not lead to recurrent waves.

\section{Interplay Between Seasonality and Mitigation}

\begin{figure}[!ht]
    \centering
    \includegraphics[width=8.6cm]{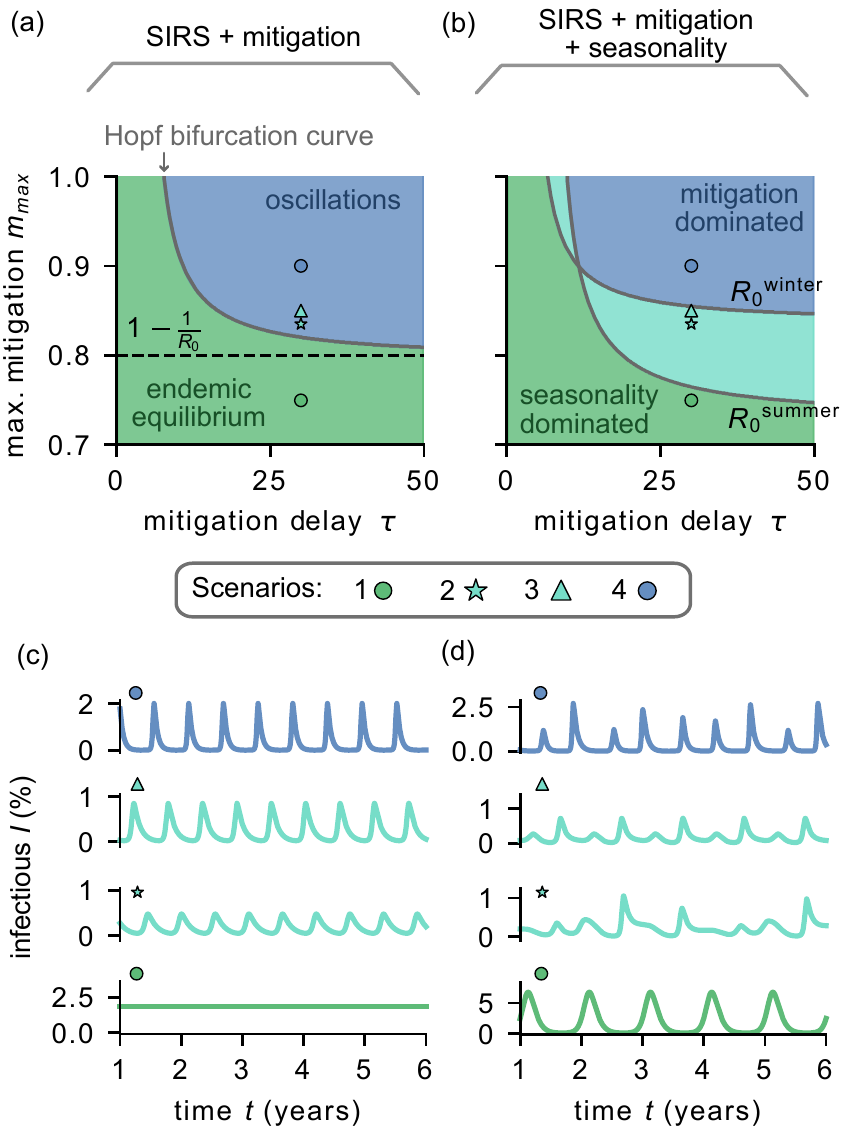}
    \caption{ 
        \textbf{The relative weighting between mitigation and seasonality gives rise to three qualitatively different dynamical regimes.} Four different scenarios are considered, that differ in their relative weighting of mitigation and seasonality (increasing maximal mitigation from scenario 1 to 4). (a) Stability diagram of the endemic equilibrium. Scenario 1 lies in the stable region (green zone), while scenarios 2-4 lie in the region of oscillations that are caused by a Hopf bifurcation if maximal mitigation $\mmax$ and delay $\tau$ are large enough ($\mmax\geq 1-\frac{1}{R_0}=0.8$, blue zone). 
        (b) Stability diagram of the endemic equilibrium for winter- and summer-adjusted $R_0$ (by fixing seasonality to its minimum and maximum values ($1\pm a$): $R_0^{\rm summer}=R_0(1-a)$, $R_0^{\rm winter}=R_0(1+a)$). The Hopf bifurcation curves separate three different regimes: Seasonality-dominated (green, stable endemic equilibrium), mitigation-dominated (blue, unstable endemic equilibrium) and interfering (cyan, stable endemic equilibrium for $R_0^{\rm winter}$ and unstable for $R_0^{\rm summer}$).
        (c) Without seasonality, scenario 1 has a stable endemic equilibrium and scenarios 2-4 display periodic oscillations. 
        (d) With seasonality, scenarios 1-4 show four qualitatively different dynamics (from bottom to top): Yearly waves of infections (scenario 1), chaotic dynamics (scenario 2), two waves per year (scenario 3) and high (or quasi-) periodic dynamics (scenario 4).
        Scenario parameters: $R_0=5$, $a=0.25$, $\tau=30$, $\mmax=0.75, 0.835, 0.85, 0.9$.
        }
    \label{fig:2-temporal-simulations}
\end{figure}

When combining seasonality and behavioral feedback the dynamics strongly depends on their relative strengths. It gives rise to three qualitatively different dynamical regimes (Fig.~\ref{fig:2-temporal-simulations}b), which are subsequently explained using four example scenarios. These represent model configurations in which mitigation is effectively weaker than (scenario 1), stronger than (scenario 4), or of comparable strength (scenarios 2 and 3) than seasonality.

Scenario 1: If maximal mitigation is too weak to break waves even in summer, namely $(1-\mmax)(1-a)R_0 >1$, then mitigation plays a secondary role. Without seasonality, mitigation is too weak to induce periodic outbreaks through a Hopf bifurcation, leading to a stable endemic equilibrium (green, Fig.~\ref{fig:2-temporal-simulations}a,c). If subject to seasonality, then only seasonal waves emerge (Fig.~\ref{fig:2-temporal-simulations}b,d).

Scenario 4: if on the contrary, maximal mitigation $\mmax$ is sufficiently strong to break waves even in winter ($(1-\mmax)(1+a)R_0 < 1$), mitigation is the dominating factor. Without seasonality, oscillations emerge through a Hopf bifurcation (blue, Fig.~\ref{fig:2-temporal-simulations}a,c). Including seasonality only affects the outbreak sizes but not so much their timing, leading to small summer waves and large winter waves. The dynamics is thus bound to an invariant torus in state space, which leads to (high-)periodic (i.e. $p$ waves in $q$ years) or quasi-periodic wave patterns (Fig.~\ref{fig:2-temporal-simulations}d, blue).

Scenario 2 and 3: in the intermediate case, seasonality and mitigation are of comparable importance. An area between the summer- and winter-adjusted Hopf bifurcation curves arises (cyan, Fig.~\ref{fig:2-temporal-simulations}b). Here, both factors interfere with each other and the dynamics is highly sensitive to small parameter changes. Scenarios 2 and 3, that only differ slightly in $\mmax$, feature qualitatively different dynamics: Scenario 3 displays two yearly waves while scenario 2 displays chaotic dynamics (Fig.~\ref{fig:2-temporal-simulations}d, triangle vs. star). 
The resulting complex dynamical regimes can be best understood when comparing the disease model to a driven nonlinear oscillator. 

\section{Emergence of Complex Infection Dynamics}

A classical example of a driven nonlinear oscillator is the driven van der Pol (vdP) oscillator \cite{van1926lxxxviii,ginoux2012van,parlitz1987period,Mettin_et_al_1993,datseris2022nonlinear}, described by

\begin{align}
            \underbrace{\ddot x + d(x^2-1)\dot x +x}_{\Omega_{\rm nat}} =& \underbrace{a \sin(\omega t)}_{\omega}\,.
            \label{eq:vanderpoloscillator}
\end{align}

$d$ represents the strength of damping and $a$ the coupling to the external sinusoidal driving. Without external driving ($a=0$), the oscillator displays nonharmonic periodic oscillations of its natural frequency $\Omega_{\rm nat}$, which depends on $d$. However, with external driving of frequency $\omega$, the resulting true frequency (if it exists, i.e., if the motion is periodic) is not the natural frequency $\Omega_{\rm nat}$ but also depends on the driving frequency $\omega$ via the coupling strength $a$. 

Similarly, in the extended SIRS model, the number of infections $I$ follows Eq.~\ref{eq:modelmaintext2}.
In the absence of seasonality ($a=0$), the system oscillates at frequency $\Omega_\tau$, which is a function of the mitigation delay $\tau$ and the other model parameters. Seasonality however, imposes an external force, stimulating the system to oscillate at the yearly frequency $\omega$. Hence, the infection waves show a resulting frequency $\Omega$ that does not only depend on $\tau$ but also on the seasonal frequency $\omega$ and its amplitude $a$, with the latter serving as the coupling to the external force. 
All in all, it can thus be expected that the extended SIRS model shows similarly complex dynamics as the driven vdP oscillator.

\begin{figure}[h]
    \centering
    \includegraphics[width=8.6cm]{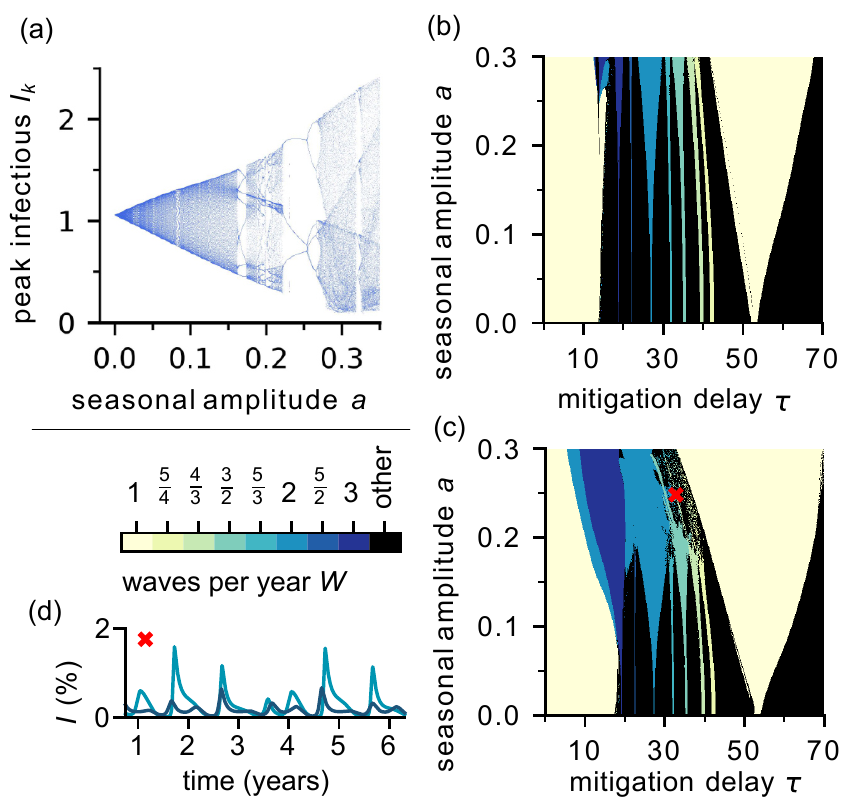}
    \caption{\textbf{Chaos, phase-locking and coexisting attractors in the extended SIRS model.} 
    (a) A \textit{peak diagram}, i.e., a scatter plot of peak infection numbers $I_k$ of the steady-state timeseries against the seasonal amplitude $a$. 
    Parameters: $\tau=32.5$, $\mmax=0.85$.
    (b,c) Arnold tongues for the averaged number of waves per year $W$ emerge as a result of phase-locking. Parameters: $a=0.25$. $\mmax=0.86$ in (b), $\mmax=0.84$ in (c). 
    (d) For $\mmax=0.84$, areas of different tongues overlap, giving rise to coexisting attractors. Parameters: $a=0.248$, $\tau=32.8$ (red cross), $S(0)=0.521$ and $I(0)=0.001$ vs. $I(0)=0.361$.
    }
    \label{fig:2regimes}
\end{figure}

The complexity of emergent infection dynamics can be visualized in a \textit{peak diagram}: By plotting the peak heights $I_k$ of a timeseries in its endemic state against a control parameter, e.g. $a$, one obtains a diagram akin to a classic orbit diagram (Fig.~\ref{fig:2regimes}a). Like in the vdP oscillator, increasing the coupling strength $a$ leads from low- and high-periodic dynamics to chaotic dynamics through period-doubling cascades \cite{olsen1990chaos,parlitz1987period}. 
For small seasonal amplitude $a$, seasonality only modulates the peak heights but does not affect the timing of infection waves. The dynamics is bound to an invariant torus in state space, which leads to phase-locked (i.e. $p$ recurring waves in $q$ years) or quasi-periodic wave patterns, depending on the ratio between the mitigation-induced frequency $\Omega_\tau$ and the seasonal frequency $\omega$. 
If seasonality has an impact comparable to that of mitigation, chaotic wave patterns can arise, as infection waves are triggered and broken by an interplay of mitigation, seasonality and immunity. 

As in the driven vdP oscillator, Arnold tongues manifest in, e.g., the averaged number of waves per year, $W$ (or winding number \cite{parlitz1987period,Mettin_et_al_1993,datseris2022nonlinear}). These represent areas of phase-locking in the $\tau$-$a$-parameter plane (Fig.~\ref{fig:2regimes}b,c). The phase-locked areas grow in size with increasing seasonal amplitude $a$, leading to tongue-like structures. 
Importantly, the shapes of the Arnold tongues depend on the maximal mitigation $\mmax$. If $\mmax$ is strong, the Arnold tongues are well separated (Fig.~\ref{fig:2regimes}b), whereas for weaker $\mmax$, the Arnold tongues start to overlap (Fig.~\ref{fig:2regimes}c). The overlapping regions feature chaotic regions, as well as coexisting attractors with very different infection dynamics, that arise for the same set of parameters but different initial conditions (Fig.~\ref{fig:2regimes}d).

\begin{figure*}[!ht]
    \centering
    \includegraphics[width=17.2cm]{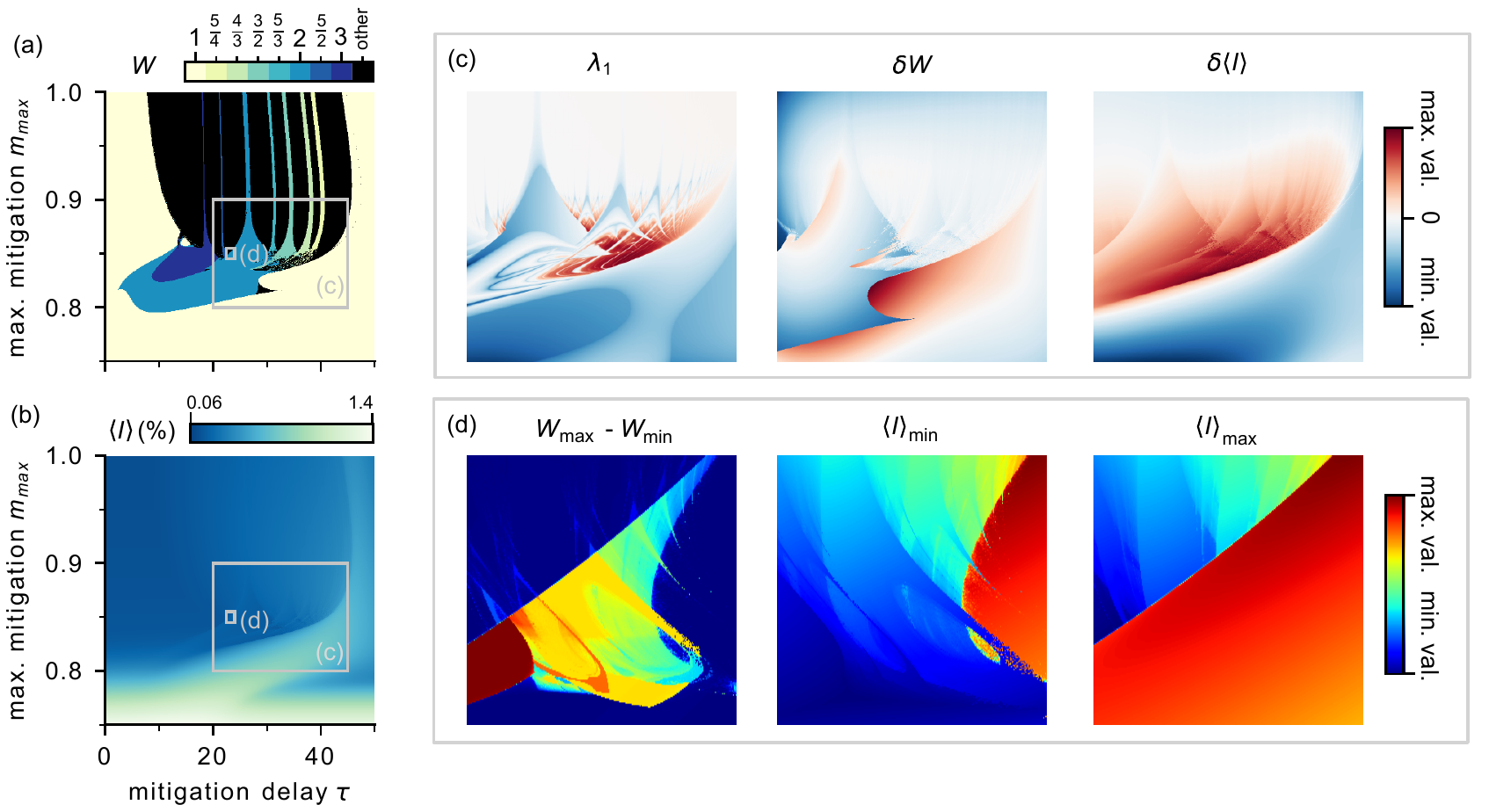}
    \caption{\textbf{Different measures reveal complex infection dynamics for various mitigation strategies.}
       (a) Arnold tongues for the number of waves per year $W$ in the $\tau$-$\mmax$-plane. 
       (b) Time averaged infections $\langle I \rangle$ drop steeply if $\mmax$ is large enough. The complexity of the dynamics in that region is characterized by different quantities visualized in two insets (gray rectangles, in panels (a),(b)): 
       (c) The largest Lyapunov exponent $\lambda_1$ is as large as $\lambda_1 = \SI{0.62}{years^{-1}}$, which corresponds to a Lyapunov time of $\SI{1.6}{years}$ (left). Parameter variations obtained by Gaussian-weighted samples of trajectories around each point (widths $\sigma_\tau=4$, $\sigma_{\mmax}=0.03$) change the number of waves per year $\delta W$ as large as $\delta W=0.53$ (absolute, mid) and the expected change of average infections $\delta \langle I \rangle$ as large as $\delta \langle I \rangle=0.27\%$ (absolute, right). 
       (d) The number of waves per year $W$ and average infections $\langle I \rangle$ differ between coexisting attractors, approached by different initial conditions. This can lead to one additional peak every two years ($\Delta W=0.5$, left) and up to $50\%$ more infections (local difference between the attractors with minimal infections $\langle I\rangle_\mathrm{min}$ (mid) and maximal infections $\langle I\rangle_\mathrm{max}$ (right).
       [min,max] values of the color bar, from left to right: (c) [-3.02,0.62 yr$^{-1}$], [-1.04,0.53], [-0.34\%,0.27\%]; (d) [0,0.5], [0.14\%,0.22\%], [0.14\%,0.22\%].
       }
    \label{fig:3-complexity}
\end{figure*}

Assuming that the seasonal amplitude is a fixed quantity (here $a=0.25$), the impact of mitigation in the model is determined by its maximal strength $\mmax$ and its delay $\tau$ (Fig.~\ref{fig:3-complexity}). As the relative strength between mitigation and seasonality is what shapes the infection dynamics, fixing the seasonal amplitude and varying maximal mitigation $\mmax$ has a similar impact as fixing $m_{\rm max}$ and varying $a$. As a result, Arnold tongues emerge in the $\tau$-$\mmax$-plane as well (Fig.~\ref{fig:3-complexity}a). They overlap where mean infection numbers increase steeply, which occurs around $\mmax \approx 1-\frac{1}{R_0} = 0.8$ (Fig.~\ref{fig:3-complexity}b). This region is characterized by complex infection dynamics, which would impede forecasting efforts in three different ways:

First, chaotic regions arise, where the largest Lyapunov exponent $\lambda_1$ is $\lambda_1 \approx \SI{0.62}{years^{-1}}$ (Sec.~\ref{subsec:largestlyapunovexponent} for details), which implies a significant divergence given slightly different initial conditions after about 1.6 years (Fig.~\ref{fig:3-complexity}c, left panel).
Second, due to the rugged parameter-space,  parameter variations ---which are expected in a realistic scenario--- can lead to abrupt jumps between different dynamical regimes. Assuming Gaussian noise (with widths $\sigma_\tau=4$ and $\sigma_{\mmax}=0.03$) around any point, the number of waves per year $W$ can change by up to $\delta W = \pm 0.5$, and the average infections $\langle I \rangle$ by $\delta \langle I \rangle=0.27\%$ (Fig.~\ref{fig:3-complexity}c, mid and right panel).
Third, the coexistence of different attractors leads to similar variations in the number of waves per year $W$ and mean infections $\langle I \rangle$, given different initial conditions (Fig.~\ref{fig:3-complexity}d). The coexistence of different attractors implies that the same mitigation behavior may lead to $50\%$ more infections, either by switching between attractors due to external perturbations, or purely due to different initial conditions.

Complex infection dynamics only occurs for a certain subset of the parameter space (Fig.~\ref{fig:3-complexity}a). For instance, it requires that the mechanisms present in the model interact on similar timescales. If the timescales mismatch, e.g., if seasonal changes are orders of magnitude slower than infection dynamics, the dynamics will cease to be complex. 

For complexity to emerge, $\tau$ must be large enough such that, without seasonality, a Hopf bifurcation leads to oscillations (Fig.~\ref{fig:2-temporal-simulations}a). However, when $\tau$ becomes too large, complex dynamics ceases to exist as the delay is so long that the mitigation kernel averages past infections over a long time period and mitigation becomes effectively weaker that way (Fig.~\ref{fig:3-complexity}a). Hence, mitigation must be slow enough ($\tau$ large enough) in order to give infection waves time to built up, and fast enough ($\tau$ small enough) to still break waves before they decay naturally due to immunity. Overall, complex dynamics is thus confined in a range of $\tau$ set by the order of the infection timescales.
Similarly, complex dynamics only arises for a certain range of maximal mitigation $\mmax$. If $\mmax$ is too weak, only seasonal waves emerge. If $\mmax$ is too strong, the dynamics is bound on an invariant torus and is high- (or quasi-) periodic. Overall, this leads to complex dynamics where $\mmax$ is just strong enough to break waves efficiently.

\begin{figure}
    \centering
    \includegraphics[width=8.6cm]{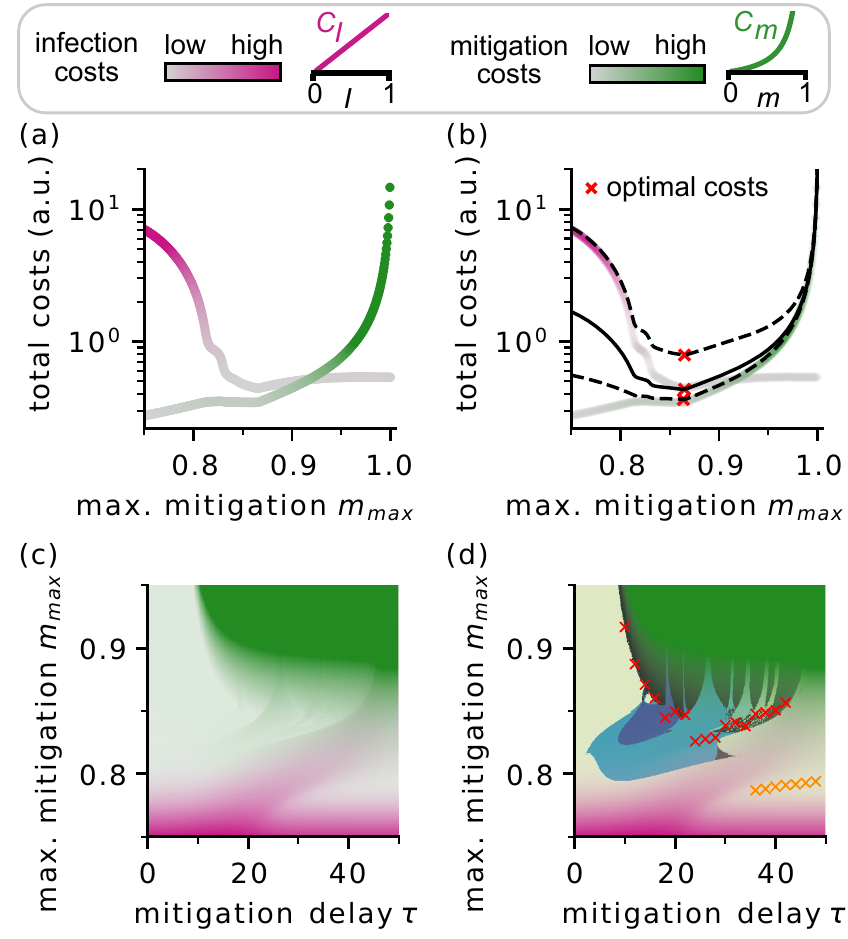}
    \caption{\textbf{Minimizing the societal costs of infections $C_I$ and mitigation $C_m$ can cause complex infection dynamics.}
    (a) Both infections and the implementation of mitigation strategies come at a cost to society. The costs increase in opposite direction for changes of maximal mitigation $\mmax$ (here $\tau=15$) (b) The cost-optimal region (red crosses on black lines) is quite invariant against changing the relative weighting parameter between  $C_I$ and  $C_m$ (dashed lines for factors of $5$ and $1/5$ relative to the solid line). (c) Costs in the $\tau$-$\mmax$-plane. 
    (d) The cost-optimal region can coincide with the region of complex dynamics, where the Arnold tongues start to overlap. Costs are overlayed onto Fig.~\ref{fig:3-complexity}a. Minimal costs are displayed as red crosses for various $\tau$. Second minima (orange crosses) emerge for large $\tau$ at lower $\mmax$. 
    The figure displays time averaged costs $\langle C_I(t)\rangle$, $\langle C_m(t) \rangle$.
    } 
    \label{fig:figure5_costs}
\end{figure}

\section{Cost Optimality of the Complex Regime}

Besides the mathematical existence of complex infection dynamics, their relevance for real world disease spread depends on whether they occur for epidemiologically important parameter regions. Both high infection numbers and strong mitigation measures come at a cost to society, creating a trade-off --- reducing mitigation costs drives up infection costs and vice versa. Hence, the mitigation employed by a society must be the cost-optimal strategy, weighing costs associated to infection numbers and mitigation. 
The following considerations will demonstrate that this cost-optimal region for mitigation may coincide with the region of complex and unpredictable infection dynamics.

A precise quantification of the multifaceted costs of infections and mitigation to society is practically impossible. We therefore take a first-order approximation for the cost of infections $C_I$ and those of mitigation $C_m$:
We assume that every single infection incurs the same expected cost, which represents all potential consequences of the infection, from sick-leave to hospitalization, death, and long-term health implications. Thus, averaged infection costs are given by $C_I = \langle C_I(t) \rangle \propto \langle I(t) \rangle$, where $\langle \cdot \rangle$ denotes a time average. They are particularly high if mitigation is not sufficiently strong to break waves on its own, namely if $(1-\mmax)R_0 > 1$ (Fig.~\ref{fig:figure5_costs}a,b). Once maximal mitigation is strong enough, the costs remain low in a wide regime.

Although favorable for reducing infection costs, implementing mitigation strategies is costly for the economy, education, culture, and individual well-being \cite{iftekhar2021look}. We assume that little mitigation is easily achievable, whereas stronger mitigation becomes increasingly more costly, e.g., because schools would need to be closed or contacts would need to be strongly reduced. Thus, we define mitigation costs as $C_m = \langle C_m(t) \rangle \propto \left\langle \left(\frac{1}{1-m(h(t))}-1\right)\right\rangle$, i.e. to be zero when the disease is not mitigated at all, and diverging if it would be completely mitigated. 

As infection and mitigation costs increase in the opposite direction for $\mmax$, a cost-optimal minimum for $\mmax$ emerges: Infection costs drop steeply for sufficiently strong $\mmax$ and remain almost flat thereafter. In contrast, mitigation costs start to increase steeply in the opposite direction. Due to the steep increase of both cost functions, the cost-optimal region for society is located where disease mitigation is "just enough" to reach the flat part of the infection costs (Fig.~\ref{fig:figure5_costs}a,b). This part coincides with the region of complex dynamics determined previously (Fig.~\ref{fig:figure5_costs}c,d and Fig.~\ref{fig:3-complexity}). Hence, by minimizing infection and mitigation costs jointly, the society might choose a maximal mitigation $\mmax$ that causes the infection dynamics to be located right in the regime of complex dynamics.

The definitions of the costs used above omit many real world effects such as nonlinearities due to hospital overload or prolonged resolute measures becoming more costly with time. Also, it is unclear how to weigh costs of mitigation against the burden of infections. Nevertheless, due to the very pronounced increase of infections at insufficient maximal mitigation, the location of the cost-optimal regime close to the sudden increase in average infections is very stable against the relative weighting of the costs (five fold change of weighting, Fig.~\ref{fig:figure5_costs}b, dashed black lines). While the costs appear lowest at small values of $\tau$ (Fig.~\ref{fig:figure5_costs}c,d), left to the complex regime, such small mitigation delay may be inaccessible to the population, due to constraints of implementing mitigation measures promptly and delays regarding the reporting of new infections.

\begin{figure*}[!ht]
    \centering
    \includegraphics[width=17.2cm]{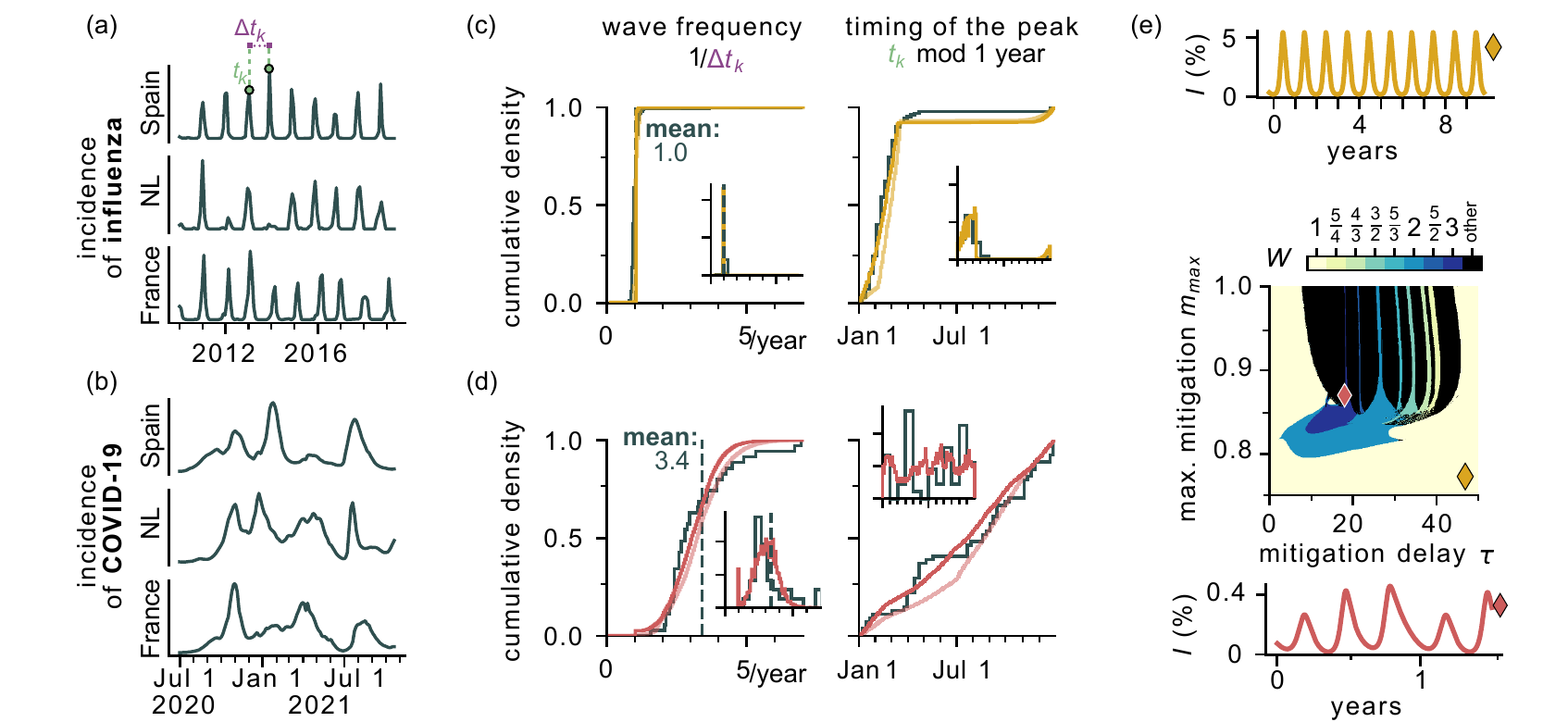}
    \caption{%
        \textbf{COVID-19 and influenza wave characteristics in northern countries are shaped by how they are mitigated respectively.} (a,b) We collected timeseries of COVID-19 (summer 2020 until the spread of the Omicron variant) and seasonal influenza (2010 to 2019) incidence in several northern countries and extracted the wave frequency $1/\Delta t_k$, i.e., the inverse time between consecutive waves, and the times of the year at which incidences peaked, $t_k$. (c) Influenza waves show a narrowly peaked distribution of wave frequencies around 1 wave per year, occurring predominantly between January and March (gray lines). (d) COVID-19 wave frequencies show a broad distribution from 1 to 7 waves per year, occurring all year long (gray lines). (e) Example model timeseries at low (yellow, top panel) and high (red, bottom panel) maximal mitigation $\mmax$ qualitatively resemble influenza and COVID-19 spread in (a,b). To account for fluctuations in real-world mitigation, we sampled model timeseries around the example points in the $\tau$-$\mmax$-plane with Gaussian weights to also extract the wave frequencies and peak timings in model simulations (widths of the Gaussian samples: $\sigma_\tau=4$, $\sigma_{\mmax}=0.03$). The resulting model distributions resemble those observed for COVID-19 in the regimes of complex dynamics (red in d,e) and those of influenza in the regime of yearly peaks and weak mitigation (yellow in c,e). Wave characteristics of the endemic (opaque lines) and the transient (semi-transparent lines) state of the model trajectories barely differ (c,d).
        }
    \label{fig:fig5data}
\end{figure*}

\section{Signs of Complex Infection Dynamics in Data}

Our theoretical results above indicate that strategies minimizing the total costs of infections and mitigation can lead societies to mitigate diseases in a way that fosters high parameter sensitivity and chaos. Can we find signatures of this in real world data? To investigate this, we analyzed the wave patterns of COVID-19 and seasonal influenza to compare them with those obtained from our model. 

Influenza and COVID-19 spread through similar channels of contagion but differ in how the human population reacted to their outbreaks. In particular, following the outbreak of COVID-19, global news coverage led to a high risk awareness among the population, causing strong mitigation measures, including school closures and lockdowns. This, combined with the influence of seasonality and fluctuating levels of immunity led to complicated wave dynamics hard to predict. The frequency of waves as well as their respective timing within the year show a broad distribution (Fig.~\ref{fig:fig5data}b,d). In contrast, while influenza is also subject to mitigation measures, such as pre-emptive vaccinations, human risk awareness is undoubtedly lower than it was for COVID-19. As a consequence, the timing of influenza waves is strongly subject to seasonality, with usually one yearly outbreak occurring predominately at the beginning of the year (Fig.~\ref{fig:fig5data}a,c). 

When comparing the empirical data quantitatively with our model, we have to acknowledge that the spread of influenza and COVID-19 did not occur in a "stationary" society, akin to a model with a fixed parameter set. To account for such parameter variability, we chose to obtain the model characteristics (wave frequency and timing) from a Gaussian-weighted sample of trajectories (with widths of $\sigma_\tau=4$ and $\sigma_{\mmax}=0.03$) in the $\tau$-$\mmax$-plane (Fig.~\ref{fig:fig5data}e middle panel). It shows influenza-like dynamics (a single yearly infection wave between January and March) when seasonality is stronger than mitigation (small $\mmax<1-\frac{1}{R_0}\approx0.8$, Fig.~\ref{fig:fig5data}c yellow, e top panel). Importantly, broad distributions of the number of yearly waves and their respective timing, as for COVID-19, are best reflected by the complex regime of our model where diverse dynamical regimes co-exist (moderate $\mmax$, Fig.~\ref{fig:fig5data}d red, e bottom panel, see Fig.~\ref{fig:A2-data-extended} for distributions in the full $\tau$-$\mmax$-plane).

The analysis conducted throughout this manuscript analyzed the steady state of the model, where concepts such as chaos are well defined. This is not the case for the data we compared the model to. For COVID-19, the data collected followed the initial outbreak, and also for influenza it is debatable whether the disease can be considered to have reached an endemic steady state \cite{cohen2022projecting}.  
We thus obtained the wave characteristics of the model separately for both the endemic steady state as well as the transient phase after the initial outbreak (Fig.~\ref{fig:fig5data}c,d opaque vs. semi-transparent lines). Despite this difference, the distributions of wave frequencies and wave timings look very similar for the two phases of infection dynamics. This showcases that the initial phase displays similarly complicated infection dynamics as the endemic steady state. Overall, this comparison of empirical and model dynamics indicates that the difference in infection dynamics between COVID-19 and influenza can be attributed to the difference in how the population reacted to their outbreaks --- for the case of COVID-19 favoring mitigation that led to a highly complex region of (real-world) parameter space.

\textit{Final Remarks}---

SIR models with waning immunity or vital dynamics alone but no behavioral feedback only display damped, non-persistent oscillations around their endemic equilibrium \cite{Stirzaker1975APM,dietz1976incidence}. Yet, they can also be driven into regimes of chaos and coexisting attractors by seasonality \cite{engbert1994chance, aron1984seasonality, schwartz1985multiple, london1973recurrent} or time-varying vital rates \cite{earn2000simple}. However, can these complex long-term dynamics prevail in epidemiologically relevant parameter scenarios? Importantly, we showed that, when including delayed \textit{human behavior} \cite{d2009information}, such complex infection dynamics can be a characteristic of 'optimal' scenarios.
Specifically, complexity emerges when the population finds the cost-optimal trade-off between reacting resolutely enough to dampen disease spread, but not stronger than economically feasible (Fig.~\ref{fig:figure5_costs}). Our results suggest that optimal disease control may go hand in hand with the sacrifice of predictability of infection dynamics.

While the model has deliberately been kept simple, this mechanism is general: Complexity arises for different functional shapes of mitigation $m(h)$, and for diseases with different reproduction numbers or waning rates (Sec.~\ref{sec:s-sensitivity}). For complex dynamics to arise, mitigation has to be strong enough to break waves and its delay long enough such that infection waves have sufficient time to build up, but fast enough to still dampen the waves before their natural decay.

Similar results can be expected for diseases that spread on slower timescales if the mitigation manifests in a slow, reactive distribution of vaccination \cite{buonomo2020effects}. Planned, preemptive vaccination or mitigation, however, should merely lower the effective reproduction number. Future works optimizing preemptive mitigation, and expanding the model on realistic network topologies can further elucidate the entangled spread of the disease and information about them.

\section*{Code availability}
All code to reproduce the analysis and figures is available online on GitHub \url{https://github.com/Priesemann-Group/complexdynamics}.

\section*{Acknowledgments}

All authors received support from the Max Planck Society. SC, JW, SB and VP acknowledge funding from BMBF, RESPINOW (Project 031L0298) and infoXpand (Project 031L0300A) consortia. 
VP received funding from the Deutsche Forschungsgemeinschaft (DFG, German Research Foundation) under Germany’s Excellence Strategy-EXC 2067/390729940.

\section*{Author contributions}
Conceptualization: V.P., S.B., J.W., S.C.
Methodology: J.W., S.B., U.P., L.F.
Software: L.F., S.B., J.W.
Validation: all authors.
Formal Analysis: all authors.
Writing—Original Draft: all authors.
Writing—Review and Editing: all authors.
Visualization: S.B., J.W.
Supervision: V.P., U.P.

\section*{Competing interests}

All authors declare no competing interests.

\appendix

\begin{table*}[ht!]
\caption{\textbf{Model parameters and functions.}}
\label{tab:compartmentsfunctions}
\begin{tabular}{lp{6cm}l}\toprule
\makecell[l]{Variable} & Meaning   & \makecell[l]{Default value / Definition} \\\midrule
$\beta_0$ & Basic reproduction rate& \SI{0.5}{days^{-1}} \\
$\gamma$ & Recovery rate & \SI{0.1}{days^{-1}}\\
$\nu$ & Rate of waning immunity & \SI{0.01}{days^{-1}}\\
$\tau$ & Mitigation delay & Varied \\
$\mmax $ & Maximal mitigation & Varied \\
$\hthres$ & Threshold hazard & $10^{-3}$\\
$a$ & Seasonal amplitude & Varied \\
$\omega$ & Frequency of yearly seasonal variation &  $\frac{2\pi}{360}\,\SI{}{days^{-1}}$\\
$\epsilon$ & Feedback curvature & $1/4000$\\
\midrule 
$h$ & Perceived hazard by society &  $h(t) = \int_{-\infty}^t I(t')\, K(t-t')\, \text{d}t' $ \\
$K$ & Delaying kernel & $K(t) = \frac{1}{\tau^2}te^{-t/\tau }$ \\
$m(h) $ & Mitigation due to human behavior & $m(h) = \mmax - \frac{\mmax}{\hthres}\epsilon \log \left( 1+ \exp \left( \frac{1}{\epsilon} \left(\hthres - h \right) \right) \right) $\\
$\Gamma$ & Seasonality & $\Gamma(t) = 1 +a\cos(\omega t)$ \\
$C_I$ & Infection costs &  $C_I = \langle I(t) \rangle$ \\
$C_m$ & Mitigation costs & $C_m = \left\langle \left(\frac{1}{1-m(h(t))}-1\right)\right\rangle$ \\
\bottomrule
\end{tabular}%
\end{table*}

\section{Dynamical regimes}
\label{sec:sup-dynamicalregimes}

\subsection{Stability of the endemic equilibrium}
\label{sec:s-stability-ee}

By choosing the mitigation kernel $K(t)$ to be an Erlang kernel of second order, $K(t)=(\frac{1}{\tau})^2te^{-t/\tau }$, one can reduce the set of integro-differential equations for $S$, $I$ and $R$ to a set of ordinary differential equations \cite{smith2011introduction} for $S$, $I$, $R$, $h'$ and $h$, where $h'$ is an auxiliary compartment and $h$ is the hazard. This adds the two differential equations 

\begin{equation}
    \dot h' = \frac{1}{\tau}(I-h')\, \hspace{0.5cm } \text{and}\hspace{0.5cm} \dot h =  \frac{1}{\tau}(h'-h)\,.\label{eq:model-linear-chain-trick}
\end{equation}

Having transformed the set of integro-differential equations to a system of ordinary differential equations, the endemic equilibrium (EE) of the system is the fixed point $\text{EE}=\{S^*,I^*,R^*,h'^*,h^*\}$ defined by $\dot S = \dot I = \dot R = \dot h' = \dot h = 0$. In order to study the stability of the endemic equilibrium, the seasonal forcing $\Gamma(t)$ needs to be set to $\Gamma(t) \equiv 1$ (no seasonal amplitude, $a=0$), otherwise the system does not possess a fixed point. The endemic equilibrium is then numerically computable via 

\begin{align}
    h'^* &= h^* = I^*\,\\
    I^* & = \frac{\nu}{\gamma+\nu}\left(1-\frac{\gamma}{\beta(I^*)}\right)\,,\label{eq:computationEE1}\\
    S^* & = \frac{\gamma}{\beta(I^*)}\,,\label{eq:computationEE2}\\
    R^* & = \frac{\gamma}{\nu} I^*\,,\label{eq:computationEE3}
\end{align}

where we wrote $\beta(I^*) = \beta_0(1-m(I^*))$.

Its stability can be computed by evaluating the Jacobian $J$ of the system at the endemic equilibrium:

\begin{equation}
    J^* = \begin{pmatrix}
-\beta(I^*)I^*-\nu  & -\beta(I^*)S^*-\nu & 0 & -\beta'(I^*)S^*I^*\\
\beta(I^*)I^* & \beta(I^*)S^*-\gamma & 0 & \beta'(I^*)S^*I^* \\
0 & \frac{1}{\tau} & -\frac{1}{\tau} & 0 \\
0 & 0 & \frac{1}{\tau} & -\frac{1}{\tau} \nonumber\label{eq:matrixstability}
\end{pmatrix}
\end{equation}

If there is an eigenvalue with positive real part, the endemic equilibrium is unstable; if not, it is stable. In the case of no delay, namely $h=I$, the endemic equilibrium is always stable, because $m'(I)>0$ and hence $\beta'(I) \leq 0$ \cite{liu1986influence}.

\subsection{Largest Lyapunov Exponent}
\label{subsec:largestlyapunovexponent}

The extended SIRS model is chaotic when the largest Lyapunov exponent $\lambda_1$ is larger than zero. We compute it by evolving a system state $\vec{x}(t) = (S, I, h, h')$ and a perturbation vector $\vec \delta(t)$ in parallel \cite{datseris2022nonlinear}. The system state is evolved with the set of equations of the SIRS model (Eq.~\eqref{eq:modelmaintext1}, \eqref{eq:modelmaintext2} and Eq.~\ref{eq:model-linear-chain-trick}), whereas the perturbation is evolved according to the linearized set of equations:

\begin{equation*}
    \frac{\text{d}}{\text{d}t} \vec  \delta (t) = J_{\vec{x}(t)} \vec \delta (t) \,.\label{eq:llelinearised}
\end{equation*}

Here, $J_{\vec{x}(t)}$ is the Jacobian of the system at the state $\vec{x}(t)$ given by

\begin{equation}
    J_{\vec{x}(t)} = \begin{pmatrix}
-\beta(h,t)I-\nu  & -\beta(h,t)S-\nu & 0 & -\beta'(h,t)SI\\
\beta(h,t)I & \beta(h,t)S-\gamma & 0 & \beta'(h,t)SI \\
0 & \frac{1}{\tau} & -\frac{1}{\tau} & 0 \\
0 & 0 & \frac{1}{\tau} & -\frac{1}{\tau} \nonumber\label{eq:jacobianforlle}
\end{pmatrix}\,,
\end{equation}

where $\beta(h,t)=\beta_0(1-m(h))\Gamma(t)$ and $\beta'(h,t)=\frac{\text{d}}{\text{d}h}\beta(h,t)$. 
In practice, the system is evolved for 100 years until the trajectory has settled onto the attractor. Then the system (together with the linearized set of equations) is evolved for 50 years, the perturbation vector is rescaled to its original size, and the largest Lyapunov exponent is calculated via $\lambda_1 = \frac{1}{t_{\mathrm{tot}}}\log(\frac{|\vec \delta (t)|}{\delta_0})$. This procedure is repeated 40 times and the final estimate for the largest Lyapunov exponent is obtained by averaging the individual ones \cite{datseris2022nonlinear}.

\subsection{Coexisting attractors}
\label{subsec:coexistingattractors}

Depending on the initial conditions, the dynamics of the extended SIRS model can settle into different asymptotic states. Those can differ in one or multiple characteristics of the asymptotic dynamics, e.g., by a different number of waves per year $W$ or average infections $\langle I \rangle$ (Fig.~\ref{fig:3-complexity}). 

In order to find coexisting attractors, we solved the differential equations of the model for a two-dimensional grid of different initial conditions. In practice, we chose $S(0)$ and $I(0)$ in the domains $S(0)\in \left(0,1\right)$ and $I(0)\in \left(0,S\left(0\right)\right)$ for a total of 100 initial conditions. The other initial conditions were set to $R(0)=1-S(0)-I(0)$ and $h'(0) = h(0) = 0$. Even though this method does not scan the whole state space for the existence of coexisting attractors, the choices of initial conditions represent a grid of all physical starting conditions, without any hazard at the outbreak of the disease. Note that the exact number of coexisting attactors can not be estimated using this method, but it is sufficient to reveal the existence of at least two coexisting attractors by quantifying differences in the dynamics such as $W$ or $\langle I \rangle$. $\langle I\rangle_\mathrm{min}$ and $\langle I\rangle_\mathrm{max}$ in Fig.~\ref{fig:3-complexity} refer to the minimal and maximal average number of infections found between coexisting attractors for a single set of parameters. $\Delta W$ is the difference between the minimal and maximal number of waves per year between coexisting attractors.

\section{Model comparison to Influenza and COVID-19 data}
\label{sec:data-gathering}

\begin{figure*}
\centering
    \includegraphics[width=17.2cm]{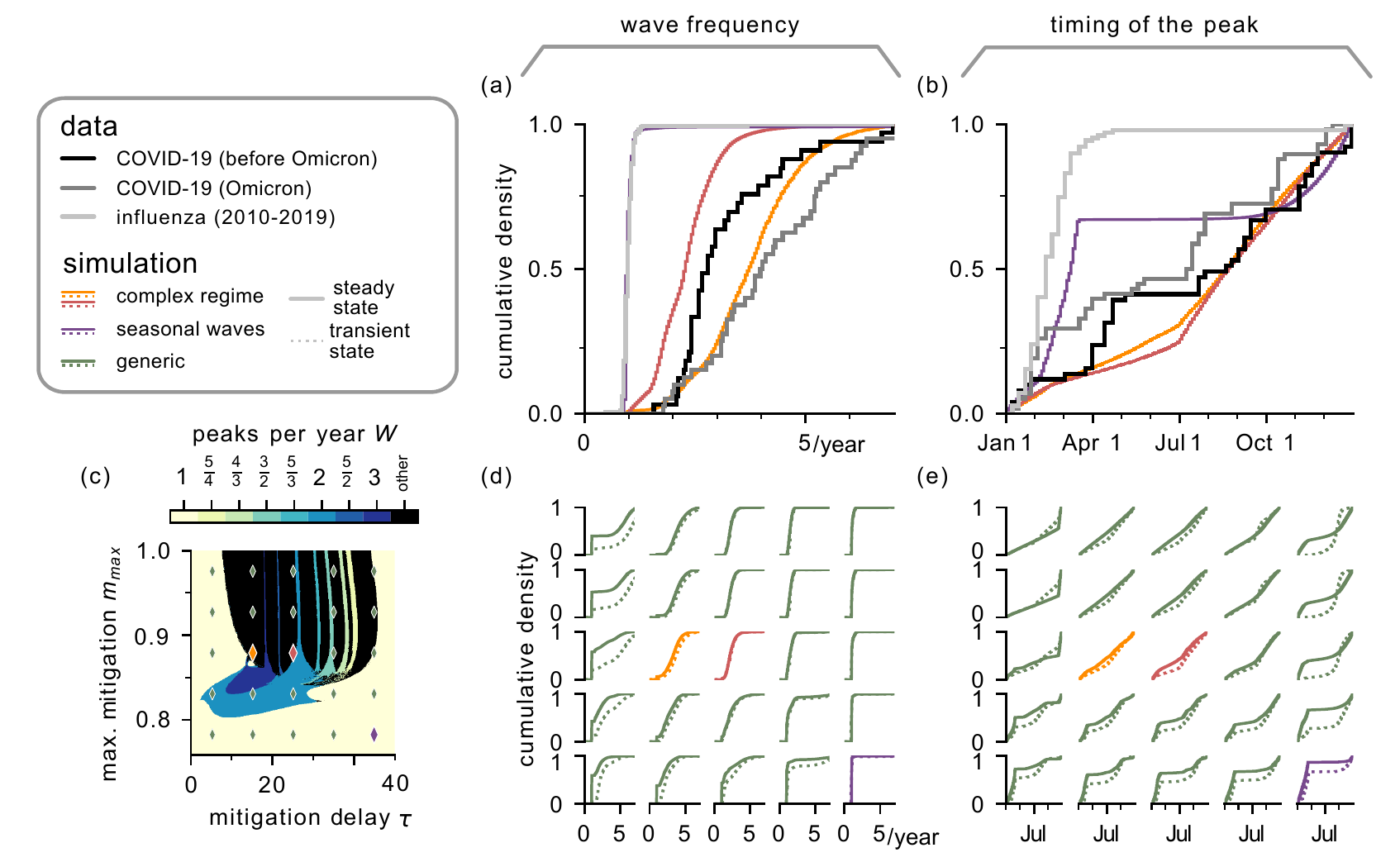}
    \caption{
    \textbf{Wave characteristics of COVID-19, influenza and simulation data.} 
    In Fig.~\ref{fig:fig5data} we compared data from COVID-19 and influenza to simulation results. Here we provide the wave characteristics for the full $\tau$-$\mmax$ parameter plane, and additionally compare COVID-19 dynamics before and after the emergence of the Omicron variant. (a), (b) COVID-19 spread of the Omicron variant (gray curves) has a slightly different distribution of wave frequencies and wave timings than pre-Omicron COVID-19 (black curves), but still resembles the distributions of model simulations in the complex regime (orange and red curves). (c)-(e) Wave characteristics in the full $\tau$-$\mmax$ parameter plane. A 5x5 grid is placed onto the $\tau$-$\mmax$-plane and Gaussian distributions of parameters are sampled around the center points of the grid (with widths $\sigma_{\tau}=4$ and $\sigma_{\mmax} = 0.03$). }
     \label{fig:A2-data-extended}
\end{figure*}

To obtain cumulative distributions for COVID-19 (Fig.~\ref{fig:fig5data}) we took the daily new confirmed cases (smoothed by a 7 day rolling average) from \textit{Our World in Data} \cite{owidcoronavirus} for 18 northern countries (Tab.~\ref{tab:datasources}). We only considered the waves from summer 2020 (July 1, 2020) up to the first waves caused by the Omicron variant (with the cut chosen to be October 20, 2021). This way, we left out the first waves in spring 2020, since they marked the initial outbreak, as well as waves caused by the Omicron variant, since it marked a considerable change in both disease dynamics and mitigation efforts. The period when Omicron dominated displayed different but qualitatively similar characteristics (Fig.~\ref{fig:A2-data-extended}). Ideally, we would have cut out the phase of mass-vaccinations in most of the considered countries in 2021 as well. However, this would have presented a drastic reduction in the size of our data set. Further, even in the first months after large-scale immunization behavioral changes adapted slowly and many mitigation efforts remained. 

We obtained the wave peaks via the \textit{find\_peaks\_cwt} method in the \textit{scipy.signal} package of Python 3.7. with 25-50 days set as the expected width of the peaks of interest. Since the algorithm sometimes returned slightly shifted peaks compared to a manual bare-eye check, we took the returned peak positions and corrected them by extracting the position of the maximum of the 40 data points in the immediate vicinity. Furthermore, we considered two peaks within 1 month as being the same up to infection and reporting noise in the data. The distributions of wave characteristics were finally obtained from an ensemble average of all countries considered.

To obtain cumulative distributions for seasonal influenza (Fig.~\ref{fig:fig5data}), we took sentinel detection timeseries from 2010 to 2019 in 24 European countries (Tab.~\ref{tab:datasources}) with weekly resolution from the Surveillance Atlas of Infectious Diseases of the European Centre for Disease Prevention and Control (ECDC) \cite{ecdc_surveillanceatlas}. This way, we filtered out early years with lacking data and years in which the spread of Influenza was significantly affected by the mitigation efforts to contain COVID-19. We obtained the peaks using the same procedure as for COVID-19 with 5-6 weeks set as the expected width of the peaks of interest, corrections of the peaks within the 6 data points in the immediate vicinity, and considered two peaks within 10 weeks to be the same up to noise.

In the simulations, we took ensembles of trajectories in the $\tau$-$\mmax$-plane, centered around different base values in this two dimensional parameter space. For each trajectory we computed the wave frequencies and timing of the waves in the steady state. We computed the cumulative distributions, weighing the trajectories with a multivariate Gaussian distribution with standard deviations of $\sigma_\tau=4$ and $\sigma_{\mmax}=0.03$. Distributions for a grid of parameter values for $\mmax$ and $\tau$ are depicted in Fig.~\ref{fig:A2-data-extended}d,e.

\begin{table*}[htp]
\caption{\textbf{Data sources for Influenza and COVID-19 timeseries}}\label{tab:datasources}
\centering
\begin{tabular}{p{2cm}p{3cm}p{8cm}p{1cm}}\toprule
\makecell[l]{} & \makecell[l]{Timeframe} & \makecell[l]{Countries} & \makecell[l]{Source} \\\midrule
COVID-19 & Summer 2020 (July 1, 2020) to before the first Omicron waves (October 20, 2021) & Italy, Spain, Portugal, Israel, Germany, France, USA, Canada, Switzerland, Austria, Netherlands, Belgium, Poland, Czechia, Croatia, Denmark, Sweden and Lithuania (18 in total)& \cite{owidcoronavirus} \\
Seasonal influenza & 2010--2019 & Austria, Belgium, Bulgaria, Czechia, Germany, Denmark, Estonia, Greece, Spain, France, Hungary, Ireland, Lithuania, Latvia, Netherlands, Norway, Poland, Portugal, Romania, Sweden, Slovenia, Slovakia, Italy, Finland (24 in total) & \cite{ecdc_surveillanceatlas}\\
\bottomrule
\end{tabular}%
\end{table*}

\section{Generality of complex infection dynamics}
\label{sec:s-sensitivity}

The model parameters used throughout the analysis in the main text have been set to match a disease with fast dynamics: A high reproduction number $R_0=5$ leads to exponential growth, mitigated at small hazard threshold $\hthres=0.001$ and the expected duration of immunity is only $\SI{100}{days}$. 
This section is intended to show that complex infection dynamics due to mitigation and seasonality also occurs for diseases with a lower reproduction number, lower waning rate, and higher hazard tolerance before mitigation is employed, as well as another mitigation function $m(h)$.

\subsection{Different reproduction number}
\label{sec:s-different_reproductionrate}

\begin{figure*}[ht]
    \centering
    \includegraphics[width=17.2cm]{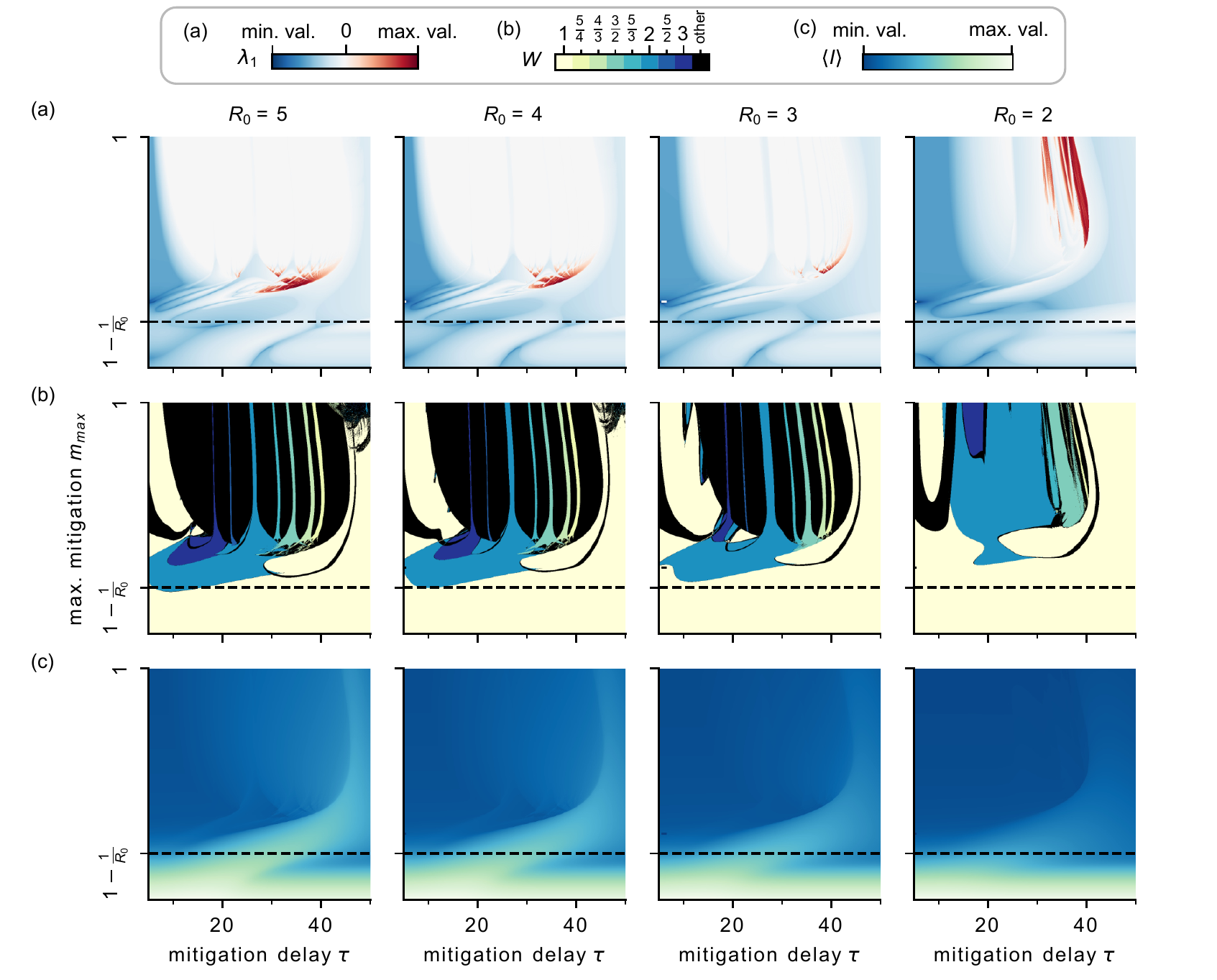}
    \caption{\textbf{Complex and chaotic infection dynamics lie at the edge of increasing mean infection numbers for different basic reproduction numbers.} The largest Lyapunov exponent $\lambda_1$ (a), average waves per year $W$ (b) and mean infections $\langle I \rangle$ (c) for different basic reproduction numbers $R_0=5$, $R_0=4$, $R_0=3$ and $R_0=2$. $R_0$ was changed by adapting the reproduction rate $\beta_0$ and keeping the recovery rate $\gamma = 0.1$ constant. Mean infection numbers increase around $\mmax \geq 1-\frac{1}{R_0}$ for all values of $R_0$. [min,max] values of the largest Lyapunov exponent $\lambda_1$ and average infections $\langle I \rangle$: $R_0=5$: $\lambda_1$:[-7.05,0.61 yr$^{-1}$], $\langle I \rangle$:[0.11\%,7.23\%]. $R_0=4$: $\lambda_1$:[-6.45,0.60 yr$^{-1}$], $\langle I \rangle$:[0.10\%,7.15\%]. $R_0=3$: $\lambda_1$:[-6.81,0.43 yr$^{-1}$], $\langle I \rangle$:[0.08\%,7.05\%]. $R_0=2$: $\lambda_1$:[-6.61,0.51 yr$^{-1}$], $\langle I \rangle$:[0.06\%,6.90\%].}
    \label{fig:s-different-r0-beta}
\end{figure*}

The default value for the basic reproduction number $R_0$ has been set to $R_0=5$ in the main text. For smaller values $R_0=4$, $R_0=3$ (by changing $\beta_0$ and keeping $\gamma$ constant) the $\tau$-$\mmax$-plane looks qualitatively similar (Fig.~\ref{fig:s-different-r0-beta}): The area where chaotic regions appear and Arnold tongues emerge is located around $\mmax = 1-\frac{1}{R_0}$: The y-axis is set such that $\mmax = 1-\frac{1}{R_0}$ is located on the same height for all values of $R_0$, which shows qualitatively similar pictures for $R_0=5,4,3$. Only for $R_0=2$ the dynamics changes: The Arnold tongues are squeezed into a more narrow range in $\tau$-direction and the region where chaotic dynamics emerge moves upward, extending all the way to $\mmax=1$ (Fig~\ref{fig:s-different-r0-beta}). 

\subsection{Different waning rate and threshold hazard}

\begin{figure}[!ht]
    \centering
    \includegraphics[width=8.6cm]{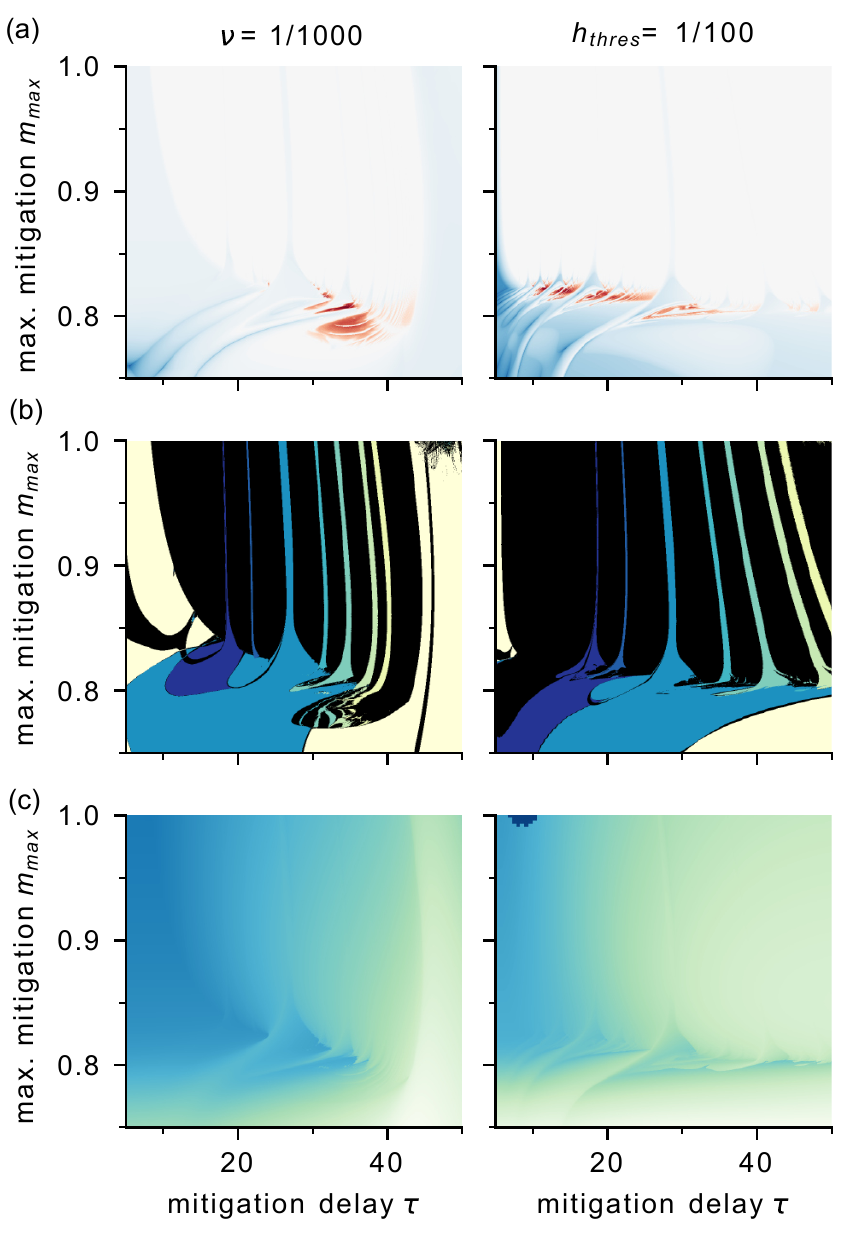}
    \caption{\textbf{Chaos and synchronization also emerge for slower waning or larger hazard threshold.} The largest Lyapunov exponent $\lambda_1$ (a), average waves per year $W$ (b) and mean infections $\langle I \rangle$ (c) for a waning rate of $\nu = 1/1000$ (10x decrease compared to default value) and a threshold hazard of $\hthres=1/100$ (10x increase compared to default value). Colors as in Fig.~\ref{fig:s-different-r0-beta}. [min,max] values of the LLE $\lambda_1$ and average infections $\langle I \rangle$: $\nu=1/1000$: $\lambda_1$:[-6.32,0.34 yr$^{-1}$], $\langle I \rangle$:[0.10\%,1.71\%]. $\hthres=1/100$: $\lambda_1$:[-6.22,0.77 yr$^{-1}$], $\langle I \rangle$:[0.86\%,9.20\%].}
    \label{fig:s-different-waning-hthres}
\end{figure}

Waning immunity is a central aspect of this study as it leads to an endemic equilibrium with $I>0$ and thus enables sustained, recurrent waves. In the main text it is fixed to $\nu = 1/100$. Together with a low threshold of hazard at which maximal mitigation is employed, $\hthres$, this makes sure that the pool of susceptible individuals never runs very low and average infection numbers are small compared to population size. If the waning rate is reduced by a factor of 10 the dynamics only changes quantitatively. Chaos and Arnold tongues still emerge, though the largest Lyapunov exponent is considerably smaller ($\lambda_1=0.34$ yr$^{-1}$, Fig.~\ref{fig:s-different-waning-hthres}). If on the other hand, the hazard threshold $\hthres$ is increased by a factor of ten, the regions of chaos in $\tau$-direction become even wider and arise for small values of $\tau$ (Fig.~\ref{fig:s-different-waning-hthres}). Altogether, the waning rate and threshold hazard must ensure that the pool of susceptible individuals is large enough to sustain Hopf bifurcation-induced oscillations. \\

\subsection*{Different mitigation feedback}
\label{sec:sup-differentfeedbacks}

\begin{figure}[!ht]
    \centering
    \includegraphics[width=8.6cm]{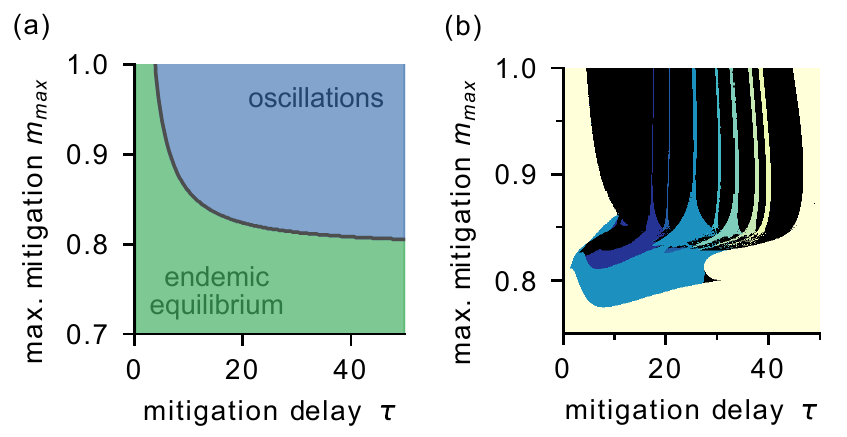}
    \caption{\textbf{Arnold tongues also emerge for a logistic mitigation feedback.} (a) Stability diagram of the endemic state for a logistic feedback $m(h)$. (b) Arnold tongues for the logistic feedback (same color scheme as Fig.~\ref{fig:s-different-r0-beta}).}
    \label{fig:s-logistic-feedback}
\end{figure}

The softplus feedback $m(h)$ (Tab.~\ref{tab:compartmentsfunctions}) used for the analysis throughout this manuscript increases linearly for small $h$ as soon as $h>0$. One could argue that a certain threshold hazard should be required in order to trigger mitigation. A suitable feedback reflecting this notion could be given by a logistic feedback with the three parameters $\mmax$, $\hthres$ and $\epsilon$: 

\begin{equation}
    m_{\rm logistic}(h) = \frac{\mmax}{1+e^{- \frac{1}{\epsilon} \left(h-\hthres \right)}}\,.\label{eq:logisticfeedback}
\end{equation}

Using this different definition, oscillations emerge again for strong maximal mitigation $\mmax$ and long mitigation delay $\tau$ (Fig.~\ref{fig:s-logistic-feedback}a). However, the Hopf bifurcation curve is shifted to the left compared to the one using the softplus feedback, implying that a shorter mitigation delay $\tau$ is sufficient to trigger oscillations. As for the softplus feedback, Arnold tongues emerge in the $\tau$-$\mmax$-plane (Fig.~\ref{fig:s-logistic-feedback}b). The same parameter values as for the softplus feedback (Tab.~\ref{tab:compartmentsfunctions}) were used.



\end{document}